\documentclass[12pt,preprint]{aastex63}
\begin{document}

%\title{Solar Chromospheric Ejections and Solar Wind Switchbacks}
\title{Solar Chromospheric Network as a Source for Solar Wind Switchbacks}

\correspondingauthor{Jeongwoo Lee}
\email{leej@njit.edu}
%\author[0000-0002-5865-7924]{Jeongwoo Lee}
\author{Jeongwoo Lee, Vasyl Yurchyshyn, Haimin Wang, Xu Yang, Wenda Cao}
\affiliation{Institute for Space Weather Sciences, New Jersey Institute of Technology, University Heights, Newark, NJ 07102-1982, USA}
\affiliation{Center for Solar-Terrestrial Research, New Jersey Institute of Technology, University Heights, Newark, NJ 07102-1982, USA}
\affiliation{Big Bear Solar Observatory, New Jersey Institute of Technology, 40386 North Shore Lane, Big Bear City, CA 92314-9672, USA}

\collaboration{1}{and}
\author{Juan Carlos Martínez Oliveros}
\affiliation{Space Sciences Laboratory, University of California, Berkeley, CA 94720-7450, USA}

\begin{abstract}
Recent studies suggest that the magnetic switchbacks (SBs) detected by the Parker Solar Probe (PSP) carry information on the scales of solar supergranulation (large scale) and granulation (medium scale). We test this claim using high-resolution H$\alpha$ images obtained with the visible spectro-polarimeters (VIS) of the Goode Solar Telescope (GST) in Big Bear Solar Observatory (BBSO). As possible solar sources, we count all the spicule-like features standing along the chromospheric networks near the coronal hole boundary visible in the H$\alpha$ blue-wing but absent in the red-wing images and measure the geometric parameters of dense sections of individual flux tubes. 
%lengths, diameters, and inter-distances between them. 
%Among these geometric quantities, the inter-distances between flux tubes appear in two scales, 0.12$^\circ$ and 1.2$^\circ$, corresponding to the medium and large scales of SBs. The length-to-diameter ratios of the dense section of flux tubes are as high as 6--40, consistent with the high aspect ratios of SBs. However, the number distributions of length and inter-distance of flux tubes appear in narrower ranges than the waiting and residence time distributions of SBs. These results suggest that the chromospheric network may produce seeds for SBs, but additional processes of aggregation, cascade, and kinking should occur during transit into space to form the SBs.
Intervals between adjacent spicules located along the chromospheric networks are found in the range of 0.4–1.5 Mm (0.03°--0.12°) tending to be smaller than the medium scale of SBs. Inter-distances between all pairs of the flux tubes are also counted and they appear in a single peak distribution around 0.7 Mm (0.06°) unlike the waiting time distribution of SBs in a scale-free single power-law form. 
Length-to-diameter ratio of the dense section of flux tubes is as high as 6--40, similar to the aspect ratio of SBs. 
Number of spicules along a network can be as high as 40--100, consistent with numerous SBs within a patch. With these numbers, it is agued that the medium scale of SBs can be understood as an equilibrium distance resulting from random walk within each diverging magnetic field funnels connected to the chromospheric networks.
%Therefore these two parameters are really important in connecting the chromosphere to the solar wind.

%These results suggest that the chromospheric network may produce seeds for SBs, but additional processes of aggregation, cascade, and kinking should occur during transit into space to form the SBs.

\end{abstract}


\section{Introduction}

Spicules occurring in the chromospheric network, also called network jets, are believed to play an important role in the transport of energy and mass into the solar wind (Withbroe 1983; De Pontieu et al. 2017; Martínez-Sykora et al. 2018), although quantitative estimates of their contribution are debatable  (Klimchuk 2012).
These small-scale ejections from the sun are of renewed interest because of the unprecedented near-sun solar observation by the Parker Solar Probe (PSP) mission (Bale et al. 2016; Fox et al. 2016; Kasper et al. 2016, Verscharen 2019). One of the most exciting observations from the PSP mission is that the near-Sun magnetic field is replete with radial magnetic field reversals called switchbacks (SBs, Bale et al. 2019; Kasper et al. 2019; Dudok de Wit et al. 2020; Mozer et al. 2020).  The observed SBs are mostly Alfv\'enic and accompanied by velocity spikes associated with the sharp increase in solar wind speed (Bale et al. 2019; Kasper et al. 2019; Dudok de Wit et al. 2020; Horbury et al. 2020; Mozer et al. 2020; Rouillard et al. 2020; Tenerani et al. 2020). Similar structures were  seen earlier from the Ulysses mission too (e.g., Kahler et al. 1996; Yamauchi et al. 2004; Suess 2007). SBs  could be tracers of the underlying source imprints in the flow, and can be energetically important for solar wind heating during expansion.

Typical scale of SBs may give a clue to their origin. Fargette et al. (2021) performed a wavelet analysis of PSP data to extract two scales that seem to bear solar origin: a large scale of 1.1$^\circ$--4.4$^\circ$ and a medium scale, 0.08$^\circ$ corresponding to 18$''$--74$''$ and 1.3$''$ on the solar surface, which can be considered as those of supergranulation (SG) and of granulation, respectively. In earlier studies based on Ulysses' observations, Yamauchi et al. (2004) interpreted that such SB size $\sim$1$^\circ$ would represent a closed magnetic loop with a network-scale separation and that the SB structure would represent an S-shaped kink formed upon interchange reconnection with neighboring open fields (cf. Fisk \& Schawadron 2001). Bale et al. (2021) found that SBs are packed within clusters in the form of intermittent magnetic structures to interprete them as in-situ remnants of magnetic funnels on SG scales. Therefore, the SG scale may not apply to one individual SB but rather to one SB patch with numerous SBs packed inside.

Shape and orientation of SBs are characterized by surprisingly high aspect ratio of the order of ten implying that SBs are long and slender along the radial direction (Horbury et al. 2020; Laker et al. 2021).  Dudok de Wit et al. (2020) investigated statistics of the waiting and the residence times (Aschwanden et al. 2016) to find them following a single power-law distribution extending over a long time range without characteristic scales. Chhiber et al. (2020) studied clustering of intermittent magnetic and flow structures using a partial-variance-of-increments (PVI) analysis; the distributions of waiting times between high PVI events also exhibits a power law at the inertial range, but followed by an exponential decrease at longer, uncorrelated lags. Both works suggest a long-term memory of SB structure, which is most likely associated with the strong spatial connection between adjacent magnetic flux tubes and their common photospheric footpoints.
Pecora et al. (2022) revisited these statistics using a larger PSP database to find the occurrence rate of SBs per unit distance fall off sharply below 0.2 au, and rise gently beyond 0.2 au.

\begin{figure*}[tbh]  % f1
\plotone{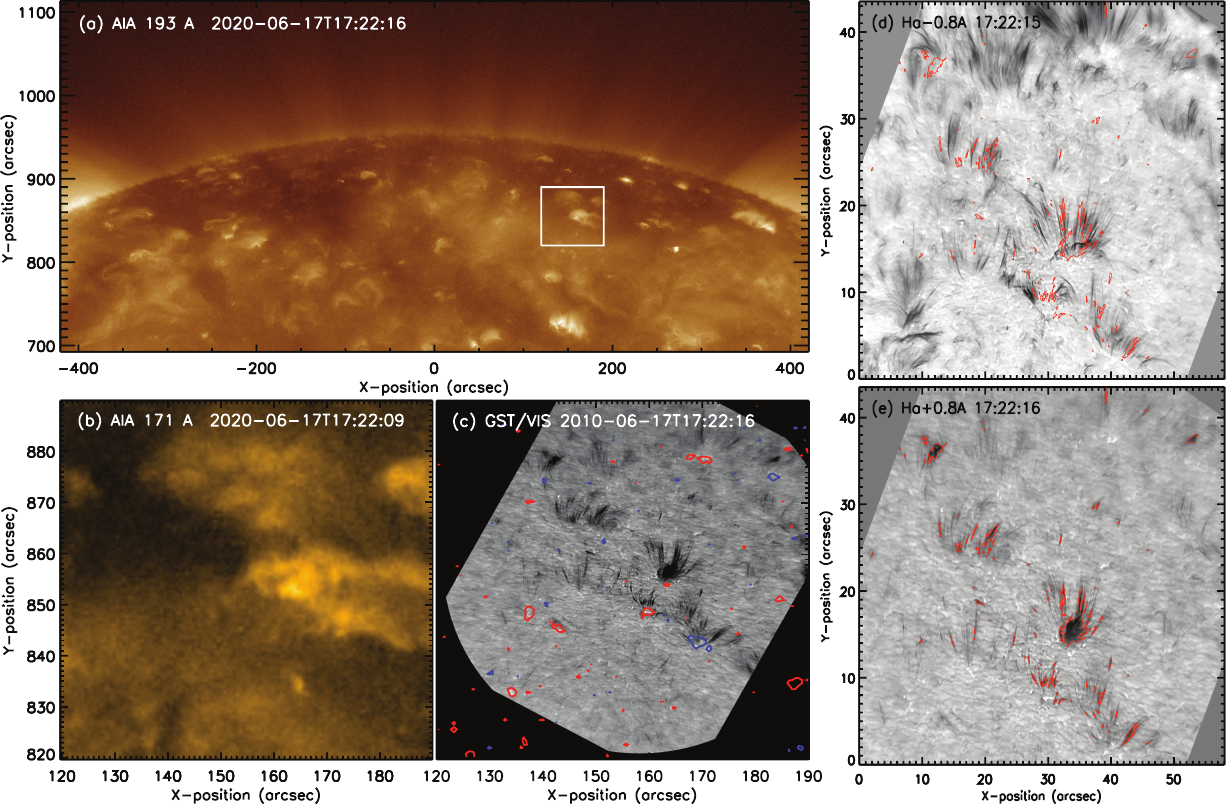}
\caption{EUV images of SDO/AIA and H$\alpha$ images of GST/VIS. (a) SDO/AIA 193 {\AA} image of the northern coronal hole on 2020 June 17 17:21:10UT, with the FOV outlined by a white box. (b) The AIA 171 {\AA} image in the ROI. (c) GST/VIS H$\alpha$ image overlaid with HMI magnetic field (red/blue contours in the levels of $\pm$30 G). (d) H$\alpha$--0.8 {\AA} image overlaid with H$\alpha$+0.8 {\AA} image (red contours). (e) H$\alpha$+0.8 {\AA} image with contours.}
\label{fig:1}
\end{figure*}

Occurrence rate of potential solar sources became an important issue for discussing the origin of SBs. Whether the sun can produce enough ejections to be detected as SBs at the time and location of PSP is yet unclear.  Sterling et al. (2020, 2016) argued for a sufficient number of small scale ejections based on the three-point distribution, citing Savcheva et al.'s (2007) rate of about 60 events per day  in polar coronal holes of size $\sim$50,000 km by 8,000 km. Nevertheless, Bale et al. (2021) claimed that SB patches must be a spatial structure on SG scales based on the pressure balance across the SB boundary, in which case the corresponding solar sources should nearly always be occurring in time. In the sun most small features under arcsec scales are so highly transient that they may appear as tracers of a spatial structure.

Among specific solar candidates for SBs, small-scale twisted magnetic structures were of interest in the setting of jet-producing mini-filament (Sterling et al. 2020). Jetlets and plumelets from embedded magnetic bipole structure are also proposed as the candidates (Raouafi \& Stenborg 2014; Raouafi, et al. 2016; Roberts et al. 2018; Uritzky et al. 2021; Kumar et al. 2022). Theoretical models for SBs also address either formation of S-shaped kink in the sun via interchange reconnection (Fisk \& Schawadron 2001) or transfer of helicity  via interchange reconnection between open and closed fields (Drake et al. 2021; Agapitov et al. 2022). On the other hand, Schwadron \& McComas (2021) proposed that a field line interacts with adjacent solar wind flows to reverse itself into an SB. Shoda et al. (2021) also propose that field lines rooted in the sun develop the kinks in the Alfv\'enic solar wind. 
Other models propose magnetized Kelvin–Helmholtz-like dynamics (Ruffolo et al. 2020) and expanding solar wind (Squire et al. 2020) for in-situ generation of SBs, in which case S-shaped kinks do not necessarily form in the sun.

In this Letter, we study solar network fields detected in the H$\alpha$ line as solar features that can be potential seeds for SBs. We investigate their length, width, and inter-distances between them to build statistical distributions of these geometrical quantities, which hopefully yields a clue to solar origion of SBs.

\begin{figure*}[tbh]  % f2
\plotone{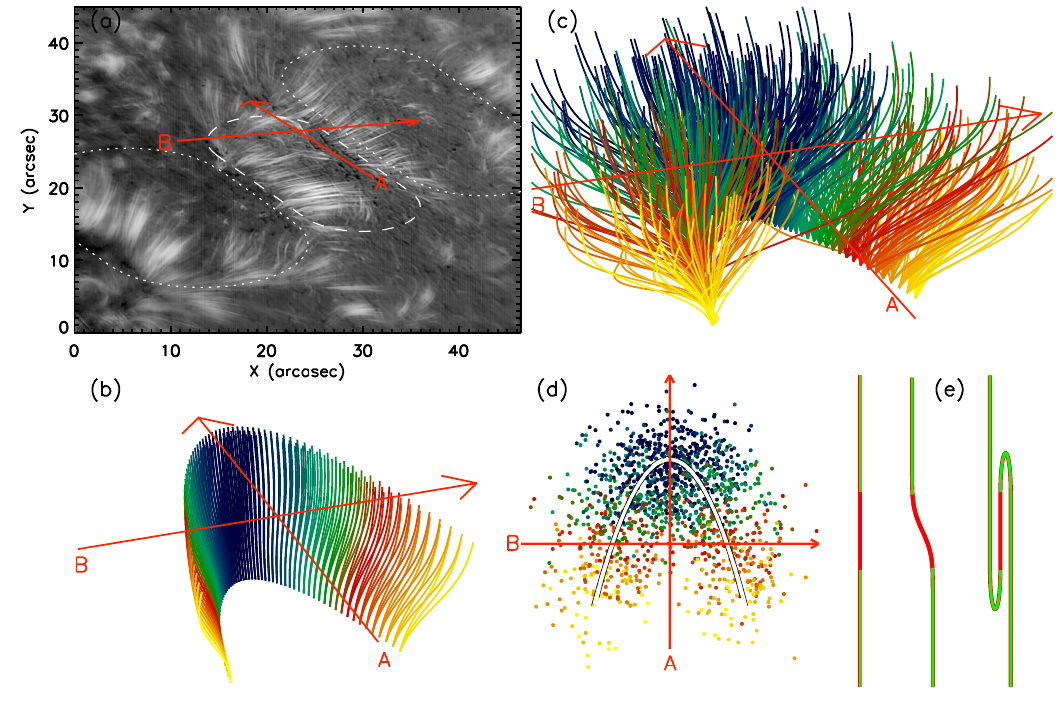}
\caption{Proposed hypothesis for the transformation of solar flux tubes into SBs in space.
(a) An inverted H$\alpha$ image showing three SGs (marked with dotted lines) and two hypothetical trajectories of PSP (red arrows).
(b) Flux tubes (represented by lines) rooted along the network boundary expand into space.
(c) Flux tubes are allowed to randomly walk to fill up the magnetic funnel yet rooted along the network boundary.
(d) Locations of flux tubes on the $x$-$y$ plane at the height of PSP are plotted (colored dots) together with the trajectory of the network boundary (white curve).
(e) Proposed transformation of a flux tube to an SB: initially a straight flux tube (green) with a dense section (red) turns itself over to form an S-shaped kink with the aspect ratio as high as that of the dense section.}
\label{fig:2}
\end{figure*}

\section{Observations}

We use the H$\alpha$ images taken by the Visible Imaging Spectrometer (VIS, Cao et al. 2010) of the 1.6 m Goode Solar Telescope (GST, Goode et al. 2010) in Big Bear Solar Observatory (BBSO). The GST observations were made on 2020-Jun-17 during the fifth perihelion period of PSP. The target region was on the northern coronal hole boundary. The GST/VIS images have a pixel size of 0$''$.029 and actual resolution is about 0$''$.10 depending on seeing. Time cadence for a set of 11 wavelengths was 28s for a fixed wavelength. All of these instrumental capabilities affect the measurement of the geometrical parameters as they are very small and highly transient.

In Figure 1 we present the location and GST field of view (FOV) of the H$\alpha$ images, and compare them with SDO/AIA (Lemen et al. 2011) EUV images. Figure \ref{fig:1}a shows the AIA EUV 193 {\AA} image including the coronal hole in the northern polar region. The white box is the FOV of panels Figure \ref{fig:1}b,c. The AIA 193 {\AA} shows a feature looking like a closed loop wrapping around the bright core (Figure \ref{fig:1}b), which does not resemble the fine structures in the H$\alpha$ image (Figure \ref{fig:1}c). Similarity between H$\alpha$ and EUV images is commonly found for active regions, but not necessarily for small H$\alpha$ features in the chromosphere.
The dark features, which we identify with spicules or flux tubes, have width of about 0.1$''$  and the inter-distance between the dark features is also in the range. Such fine structures are not clearly discernible in EUV. Those fine structures in H$\alpha$ are visible not only due to the high resolution of the GST/VIS, but an intrinsic property of the chromosphere.
Figure \ref{fig:1}c also shows HMI (Schou et al 2012) magnetogram over an H$\alpha$ image as colored contours at 40 G level. The magnetic field measurement in locations closer to the limb is harder, and from the average magnetic field in the background, we estimate that field strength lower than 40 G is not trustworthy. 
%The flux tubes lying along the network boundary are frequently erupting, and tracing them in time is also hard.

Figure \ref{fig:1}d,e shows GST/VIS images taken in the blue and red wings of the H$\alpha$ line. There is a significant red-blue asymmetry so that we can see the network structure more clearly in H$\alpha$ blue wing, and less apparent in the red wing, meaning that most materials in the network are in the upward motion. In this study we regard any thin elongated features visible in the blue wing but not in the red wing as the chromospheric ejecta.

\section{Analysis}

Prior to the analysis, we need to set a conceptual relationship between solar features in H$\alpha$ images and SBs in the PSP data, because neither S-shaped kink structure is visible in the sun nor any small-scale ejecta are traceable from the chromosphere to the position of PSP. 

\subsection{Hypotheses}
Figure \ref{fig:2} shows our proposed hypothesis on how flux tubes in the chromosphere can be related to SBs in the height of PSPs. In Figure \ref{fig:2}a we schematically mark three SG cells visible in an inverted H$\alpha$ image. Since our goal is to study fine spatial structures within an SG, the present FOV of the GST/VIS is, in fact, adequately large.  Each SG has diameter about $\sim$25$''$. In this inverted H$\alpha-$0.8 {\AA} image, the spicules appear as bright fine structures. The spicules are mostly lined up along the SG boundary with intervals of a few arcsec, and have length of arcsec and width of sub-arcsec. We do not count the closed flux tubes in the middle of the SG, because they will not reach the heliosphere. 

Figure \ref{fig:2}b shows imaginative flux tubes lining up along the SG boundary. Each line represents a flux tube extending from the chromosphere to the height of PSP flight, which is colored for identification of the footpoint location. We assume that the chromospheric structure is expanding into space in proportion to the radial distance of PSP but maintaining its original shape so that it will form a funnel with diameter of an SG (in the units of angular size) as claimed by Bale et al. (2021). If PSP flies through such a funnel along trajectory A, it will encounter only a single flux tube. Along B, PSP may encounter upto two flux tubes. Either way, PSP cannot detect SBs within a funnel as numerous as observed. 

In Figure \ref{fig:2}c, we let the flux tubes rearrange themselves to fill up the funnel via stochastic meandering (cf. Chhiber et al. 2021). Same color code is used for identifying the footpoint location of the flux tubes. In this case, PSP may encounter the SBs as numerous as observed. In Figure \ref{fig:2}d, the positions of flux tubes after the random walks (Fig. \ref{fig:2}c) are plotted with symbols and those under the simple lateral expansion (Fig. \ref{fig:2}b) are represented by the white curve. The stochastic meandering may not necessarily result in a uniform distribution of flux tubes in space, because some flux tubes may be more concentrated in some location. For instance, at a sharply turning point or in the interface between two sides of a network, PSP may find flux tubes more densely populated. Such asymmetric distribution of flux tubes within an SG funnel may result in asymmetric SB distribution, depending on the direction of PSP flight (Bale et al. 2021).

All these arguments hinge on how SBs form.  We illustrate our idea on the kinking in Figure \ref{fig:2}e. Here we assume that plasma behind the dense section (red) of the flux tube moves faster and pushes the flux tube to kink. If the dense section of the flux tube is more rigid to resist against folding itself, the folding occurs at the top and at the bottom to maintain the aspect ratio. Such kinking might also occur via Kelvin–Helmholtz instability at the SB boundaries (Mozer et al. 2020; Ruffolo et al. 2020) or under the mechanisms proposed by Shoda et al. (2021) and Swadron \& McComas (2021).
%Based on this thought experiment, we expect the following relationships between solar parameters and those found in the PSP data:
Based on this thought experiment, we set the following strategy of data analysis:

\begin{enumerate}

\item 
We count all flux tubes that transiently appear in H$\alpha$ blue wing images under the assumption that flux tubes may develop S-shaped kink during transit.

\item We measure inter-distances between flux tubes for comparison with the waiting time of SBs multiplied by the PSP speed. This actually replaces the concept of temporal occurrence rate at a given position.

\item
We measure the length of the mass-loaded section of flux tubes for comparison with the SB residence time multiplied by the speed of the ejecta.

\item
We measure the aspect ratio of the mass-loaded section for comparison with that of SBs under the assumed mechanism for S-shaped kink formation in space.

\end{enumerate}

\begin{figure*}[tbh]  % f3
\plottwo{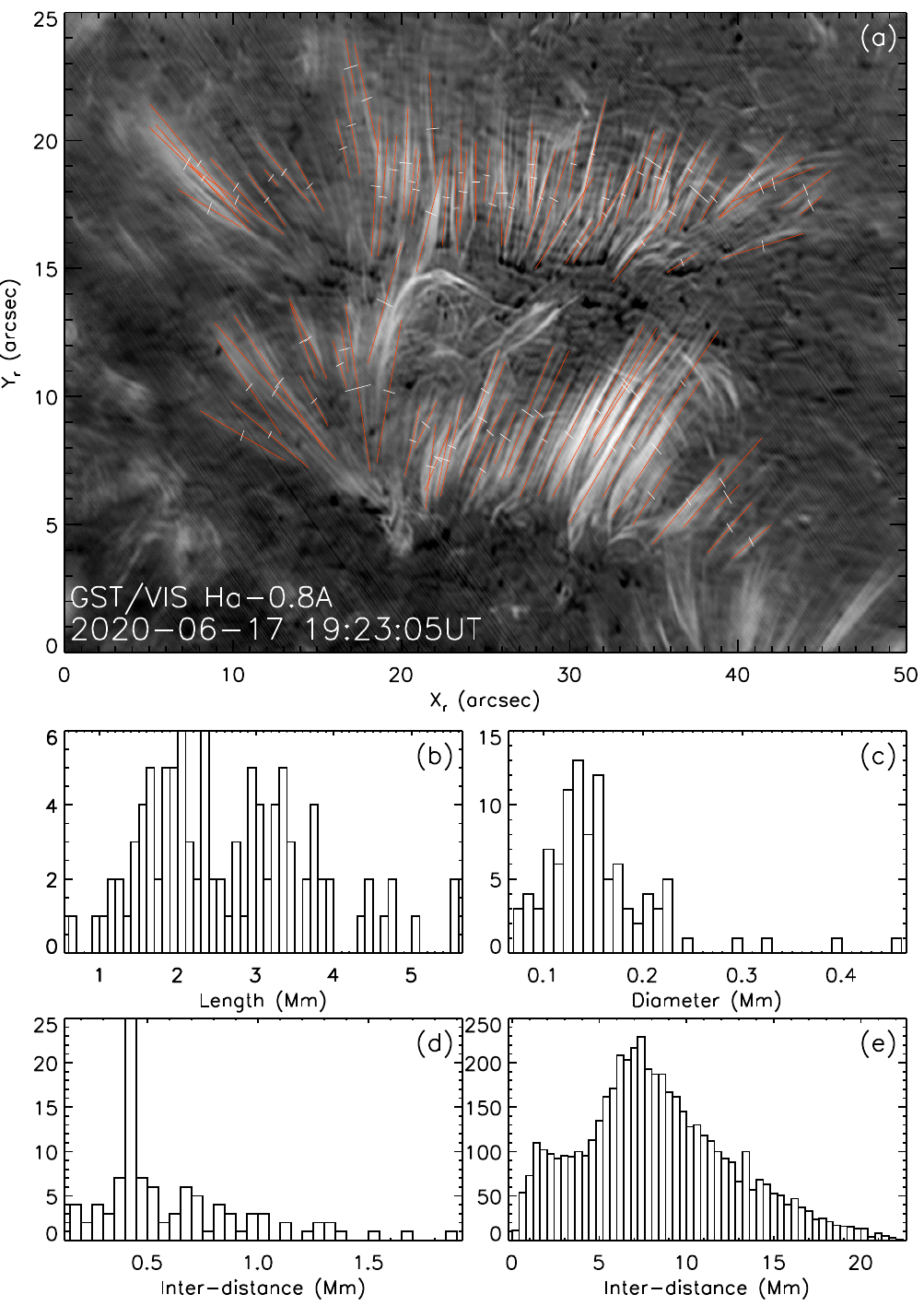}{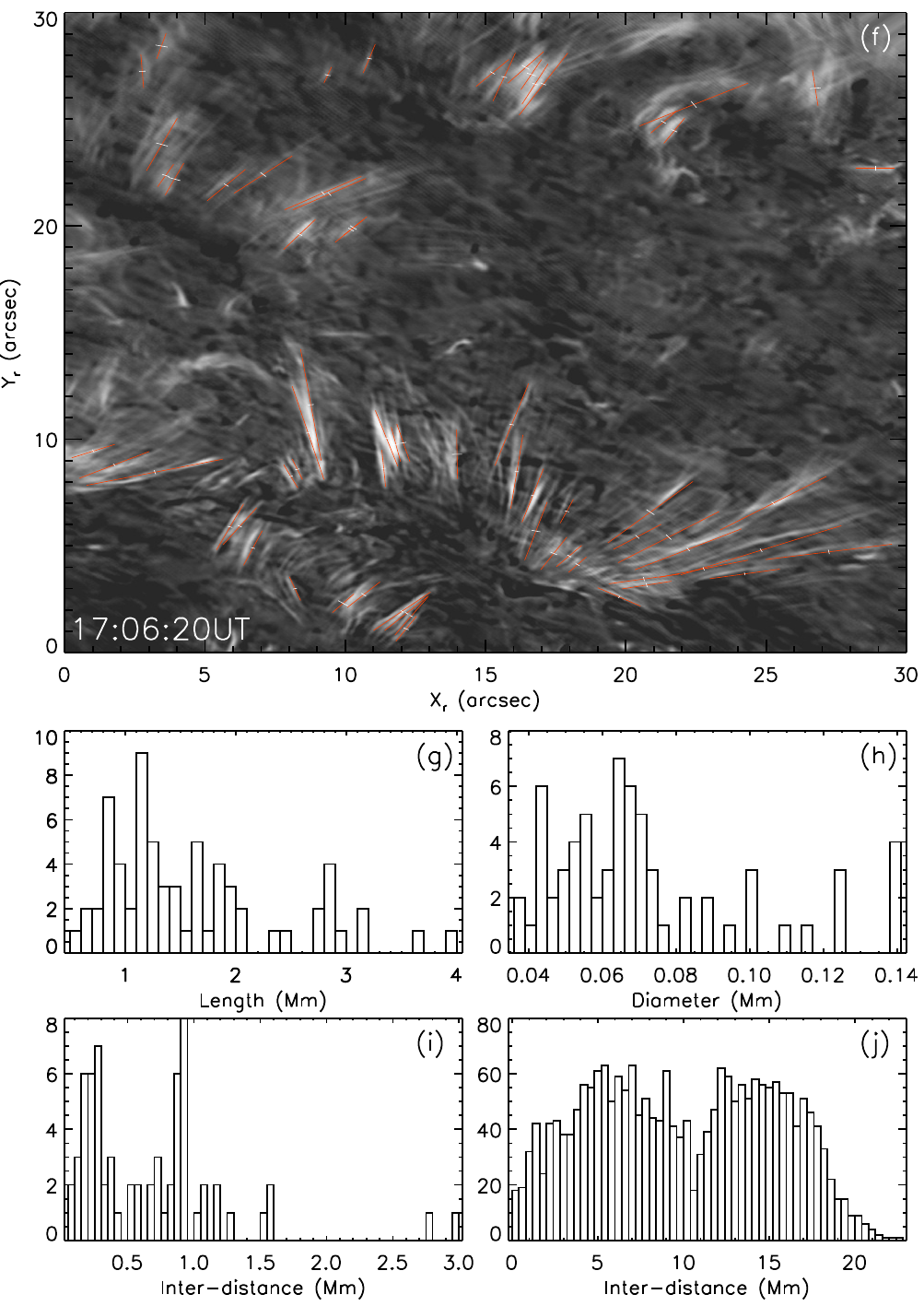}
\caption{Geometrical parameters of flux tubes on SG boundary. The panels in the left hand side show the lengths (widths) of flux tubes marked with red (white) bars in an inverted H$\alpha$ image (a),
histograms of length (b) and width (c) of flux tubes, and inter-distances between nearby flux tubes (d), and those between all pairs of flux tubes (e).
The panels in the right hand side (f--j) are the same as (a--e) for a different time frame.}
\label{fig:3}
\end{figure*}

\subsection{Individual Features}

Figure \ref{fig:3} displays the GST/VIS  images (top panels) and the measured parameters (bottom) for two contrasting frames. 
We counted only those belonging to the middle one (the dashed ellipse in Figure \ref{fig:2}a), excluding spicules belonging to neighbor networks to study characteristics of one magnetic funnel.
Closed fields inside the network are also excluded as they would not reach the heliosphere. 
In the frame shown in Figure \ref{fig:3}a, we count 100 flux tubes, which must high enough for explaining the large number of SBs within a patch. 
It is generally debatable whether spicules can reach heliosphere; of the two types of spicules, Type II spicules are believed to be more likely to escape the sun (Martinez-Sykora et al. 2018; Klimchuk 2012). We are unable to distinguish type II from type I with the current data and count all of the thin structures, which is one property of type II spicules. As a result, the total number of the flux tubes related to SBs can be overestimated.

We measure the length and diameter of the dense section of each flux tube solely based on the intensity appeared in the H$\alpha$ image, i.e., the bright fine fibrils in this inverse H$\alpha$ image. We first create a skeleton image by marking the locations of local minimum H$\alpha$ intensity. We then manually mark the top and bottom of each flux tube based on its intensity against the nearby background, after which we use an algorithm to determine its length and width.
The measured parameters are shown in Figure \ref{fig:3}(b,c) in the form of histogram. The length distribution tends to show two peaks followed by slow decays on both sides (Fig. \ref{fig:3}b). In a log-log plot they do not form a single power-law decreasing toward the larger scale unlike the statistical distribution of either waiting time or residence time (Dudok de Wit et al. 2020). The diameters are much smaller than any other parameters and form a narrow distribution below 0.5 Mm down to 0.07 Mm (Fig. \ref{fig:3}c).

Figure \ref{fig:3}d shows that the inter-distances between nearby flux tubes form a wider distribution, but it is still confined to a narrow range from 0.1 Mm to 2 Mm.
We also calculate the inter-distances for all pairs of flux tubes, as a possible counterpart to the waiting time distribution of SBs. PSP may hit any pair of flux tube lying on the edge of the magnetic funnel depending on its direction of flight. In reality, PSP will pass through a particular  funnel only once, but may encounter multiple funnels on the trajectory. The inter-distances of all pairs of flux tubes within a funnel can thus hint on the statistical behavior of the waiting times if all funnnels are more or less alike.
In this case, the total number of data points increases from $n$ to C($n$,2), where $n=100$ in Figure \ref{fig:3}a and $n=74$ in Figure \ref{fig:3}f. Two peaks are located at 1.4 Mm and 7.3 Mm as indicated by the two vertical guide lines, which correspond to  0.12$^\circ$  and 0.60$^\circ$, respectively, consistent with the medium scales found for SBs (Fargette et al. 2021).

In the right hand side of Figure \ref{fig:3}, we repeat the same procedure to another frame. The derived parameters from the image (Fig. \ref{fig:3}f) follow the above trend: the diameters are in the shortest scale and the lengths is the next. The main noticeable difference is the inter-distances from all pairs of flux tubes show two peaks clearly. The smaller one around 5 Mm corresponds to the medium scale of SBs, and the larger one around 14 Mm,to the medium scale of SBs. 
While such agreement may seem supportive of the solar origin for SBs, we must note that this SG scale is less apparent in Figure \ref{fig:3}a. It may be that the presence of numerous flux tubes (Fig. \ref{fig:3}a) leads to a continuously decreasing inter-distance distribution across the SG scale and the lesser number of flux tubes along the SG boundary produces the SG scale more clearly (Fig. \ref{fig:3}f). The latter is actually more common.
Since almost all chromospheric features at these small scales are highly transient, no time-invariant distribution could be constructed, and the result should be interpreted with caution.

Since we measured these parameters solely based on the apparent appearance on the H$\alpha$--0.8{\AA} images, the results depend on both the line opacity and projection effect. We may expect that width less suffers the projection effect, and that the inter-distance between a pair of flux tubes is free of the radiative transfer effect. However, the distance is also subject to the projection effect in case that pairs of flux tubes do not lie on the same sky plane. We might expect the ratio of the length to width be less sensitive to the radiative transfer effect, but is still subject to the projection effect. We thus expect that length, aspect ratio, and inter-distance are underestimated due to the projection effect.
Our interpretation on these quantities will be made under such limitations.

\begin{figure*}[tbh]  % f4
\plotone{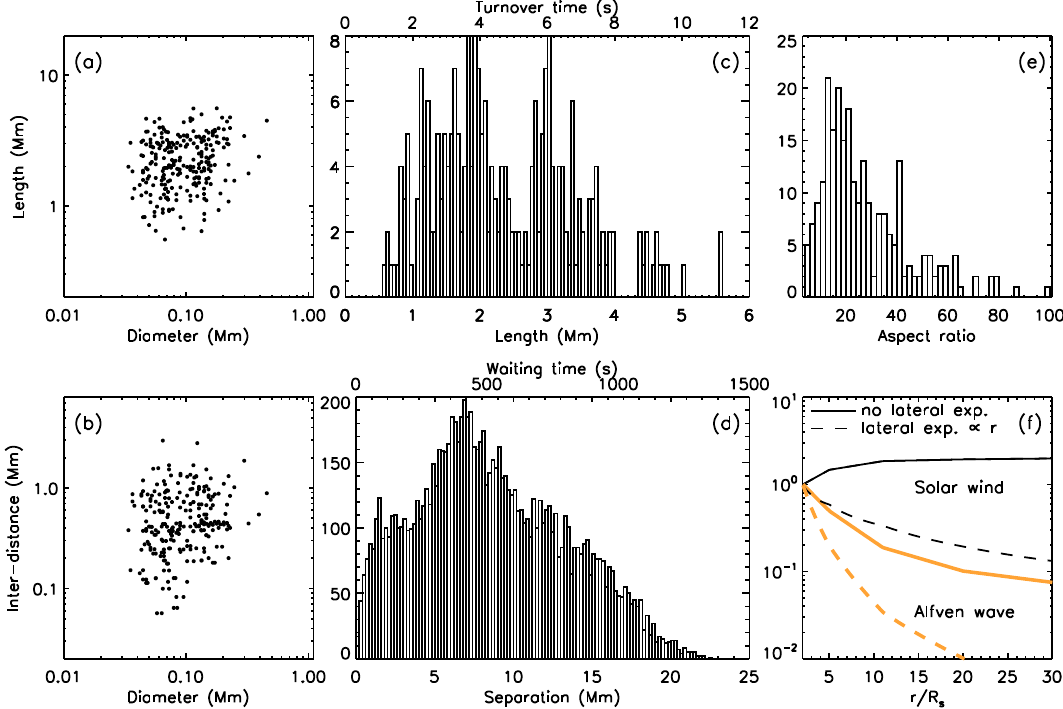}
\caption{Distributions of geometrical parameters. Scatter plots of (a) length of flux tubes vs. their widths and (b) inter-distance between nearby flux tubes vs. widths. Middle column shows the histograms of length of the dark section of flux tubes (c) and inter-distance of all pairs of flux tubes (d). Right column shows the measured length-to-diameter ratio of flux tubes (e) and a model aspect ratio (f). The model calculation of the aspect ratio starts with the nominal initial value of one in thr sun and traces its evolution through its transit to space under four combinations of assumptions on the radial and lateral expansion. }
\label{fig:4}
\end{figure*}

\subsection{Total Distributions}

We combine the results of the geometrical parameters measured from three H$\alpha$ images altogether, and plot the results in Figure \ref{fig:4}. The first column shows scatter plots of the length (Fig. \ref{fig:4}a) and the inter-distance (Fig. \ref{fig:4}b) versus the diameter where total 250 data points are used. As a result no good correlation is found either between length and diameter or between the inter-distance and the width. This means that any similarity in the number distribution of these quantities is not due to good correlation between length and width.

\begin{deluxetable*}{lllllll}
\tabletypesize{\scriptsize}\tablecolumns{6}
\tablewidth{0.95\textwidth} \tablecaption{Properties of SBs investigated\label{tab01}}
    \tablehead{
    \colhead{ } &
    \colhead{Medium scale, $d$} &  \colhead{Large scale, $D$} &
    \colhead{Aspect ratio}   &   \colhead{Number per patch, $N$}&
    \colhead{Frequency distribution} }  
\startdata
      Expected 	& 0.08$^{\circ}$ 	$^a$          & 1.6$^{\circ}$--6.2$^{\circ}$ $^a$   &                     &           &           \\
      PSP    		& 0.12$^{\circ}$--0.7$^{\circ}$ $^a$  & 1.1$^{\circ}$--4.4$^{\circ}$  $^a$  & $\geq$10  $^b$  &  -  &Broad without charcteristic scales  $^c$ \\
      GST 		& 0.04$^{\circ}$--0.12$^{\circ}$     & 0.6$^{\circ}$--1.8$^{\rm o}$  &   6--40         &  40--100  & Narrow  with charcteristic scales 
\enddata
\tablenotetext{a}{Fargette et al. 2021}
\tablenotetext{b}{Laker et al. 2021}
\tablenotetext{c}{Dudok de Wit et al. 2020}
\end{deluxetable*}

The middle column shows the lengths (Fig. \ref{fig:4}c) and the inter-distances of flux tubes (Fig. \ref{fig:4}d) in histogram.
% {as these quantities have sizes comparable to the SB scales}.
Both quantities are shown in units of Mm along with corresponding times in units of s, the latter of which are intended for a crude comparison with the residence-time and the waiting-time distributions of SBs. 
%The conversion from the distance to the time is made using a nominal speed of 500 km s$^{-1}$ and other assumptions as follow. 
%Two vertical dashed lines denote the majority of the population defined as the central part excluding top and bottom 10\%. } 
%The length of spicules mostly lie within the range of 0.5--5 Mm (marked by the red vertical dashed lines) and average within the two scales is 2.3 Mm.  show two peaks  The full range is 0.55--5.6 Mm, which corresponds to the turnover time range from 2.0 s to 10 s if this nominal speed corresponds to turnover speed of ejecta. 
The lengths of spicules lie in the range of 0.55--5.6 Mm. To convert these to turnover times, we assumed that a nominal speed of 500 km s$^{-1}$ for the ejecta, and the length of the spicule is doubled reaching the height of PSP perihelion (see the calculation result in Figure 4f). We then find the corresponding turnover times ranging from 4.0 s to 20 s, which roughly matches the width of the residence time distribution of SBs (Dudok de Wit et al. 2020). However the former distribution has two peaks, whereas the residence times form a single power-law distribution.
%the units of degree instead of Mm and appear in a single peak dstribution
%are mostly in the range of 0.1$^\circ$--1.2$^\circ$ (two vertical lines), and the peak of the distribution is located at 0.6$^{\circ}$. 
%The range between the first two scales closely reproduces the midium-scale range 0.12$^{\circ}$--0.70$^{\circ}$ of Fagette et al. (2021). The largest scale, 1.2$^\circ$, lies within the large-scale range, 1.1$^{\circ}$--4.4$^{\circ}$ (Fargette et al. 2021). However, the two scales are not clearly distinguished in this distribution, although samples from other snapshots may show such a two-peak distribution (e.g., Figure \ref{fig:3}j). 
The inter-distances lie in the range of 0.057--22 Mm, which can be translated to 3.4--1300 s if PSP runs across these structures with the nominal speed, 500 km s$^{-1}$ at the radial distance of 30 $R_\odot$. This range is close to that of the waiting time distribution of SBs, but the distribution has a peak around 7 Mm unlike the scale-free single power-law (Dudok de Wit et al. 2020).
The actual speed and radial distance will vary from these nominal values and our assumptions are too simple. Nevertheless the difference between chromospheric parameters and their solar wind counterparts is obvious.

The rightmost column is devoted to the discussion of the aspect ratio of SBs. Figure \ref{fig:4}e shows the ratios of the length to the diameter of the dark sections of the flux tubes, which we equate to the aspect ratio of a S-shaped kink, as shown in Figure \ref{fig:2}e. 
This parameter mostly lies in the range of 6--40, which well fits the range of the aspect ratio of the SBs detected by the PSP.   
Figure \ref{fig:4}f presents a simple calculation result on how aspect ratio of a flux tube varies as it propagates out with a unit aspect ratio in the sun.
This calculation is made using the Alfv\'en speed, $V_A$ and solar wind speed $V_{SW}$ presented in Adhikari et al. (2020). 
If the tube moves at $V_A$ which is decreasing with radial distance, the top part is moving more slowly than the bottom part, and the tube becomes compressed in length and thus the aspect ratio decreases regardless of the assumption on the lateral expansion.  If it moves at $V_{SW}$  which is increasing with radial distance, its length increases, and the aspect ratio may increase for invariant diameter. 
Even in this most favorable scenario, the aspect ratio increases only by a factor of 2; otherwise it only decreases with radial distance. This simple exercise implies that a flux tube should have already had a highly elongated shape at the start in order to reproduce the high aspect ratio of SBs in space. 
 
\subsection{Redistribution of Flux Tubes in Space}

We list our results in the bottom row of Table 1, along with the other results from the PSP data in upper rows, for comparison. We hoped that the chromospheric scales, when expressed in angular distance, may be comparable to the medium scale ($d$) and large scale ($D$) of SBs (Fargette et al. 2021; Bale et al. 2021). %Assuming that the chromospheric structures are somehow preserved.
%This assumption is, however, not guaranteed (Figure 2), and we wish to check to which extent this assumption is valid, and whether additional fine tuning of the quantities listed in the table can be made otherwise.
However, the chromospheric scales found in this study tend to be smaller than $d$ and $D$ determined from the PSP data in overall (Table 1).
%, and is of new interest whether additional fine tuning of those quantities listed in Table 1 can be made.
%in reference to a number of papers on the random walk of magnetic field lines in solar wind turbulence (Chhiber et al. 2021; Ruffolo et al. 2003; Tooprakai et al. 2016).
In the presence of solar wind turbulence, it is obvious that at some point far away from the sun, the chromospheric scale information will be lost as a result of random walk of field lines in solar wind.  Ruffolo et al. (2003) and Tooprakai et al. (2016) calculated the heliocentric distances up to which field lines originating in magnetic islands can remain topologically trapped and estimated this so-called filamentary distance as 40--150 R$_\odot$. The PSP data used in the scale analysis (Fargette et al. 2021) was gathered at its perihelion distance $\sim$30 R$_\odot$, 
%just below this filamentary distance. 
%This filamentation is not due to photospheric motions (Giacalone et al. 2000) but in situ interplanetary turbulence (Ruffolo et al. 2003; Zimbardo et al. 2004).
and therefore the filamentary structure could barely be maintained. 
Chhiber et al. (2021) used a turbulence transport model to estimate that the diffusive field line spreading at 1 au is about 20°--60°. A linear interpolation of this value to the height of 30$R_\odot$ yields 2.8°--8.4°, which must be an overestimation.
%Another thing to consider is that field lines emanating from an SG may be neighbored by others from another SG in which case they may not expand as freely as those from a single isolated SG.
We may still expect that size of the funnel structure at the height of PSP perihelion, expressed in angular distance, can be larger than an actual chromospheric SG (Table 1) as a result of lateral displacement and spread. 
 
Likewise we may expect that $d$ determined from the PSP data should be also larger than actual granule size. However, we further assumed the presence of random walk of magnetic flux tubes within each SG in addition to the expansion and spread, because in order for a number of the flux tubes rooted in the SG boundary to be detected by a satellite they should fill up the funnel quite significantly (Figure 2). The size of granules would be forgotten after this random walk, even though granules are involved with the flux tubes.  
%A supporting model result can be found in global simulation of interplanetary magnetic field lines show that ...the PSP-observed scales represent the transverse scale of the shuffling process (random motions + expansion, according Chhiber et al (2021), and they have little or nothing to do with the chromospheric scales (Chhiber et al. 2021).
We consider an extreme case in this line that they are completely rearranged in position so that the inter-distance between filament becomes the same for every pair of the nearest neighbors, and $d$ is expressed as $D / N^{1/2}$. With the chromospheric parameters, 0.6°$\leq D\leq$1.8° and $40\leq N\leq 100$, we find 0.06°$\leq d\leq$0.28°, which is similar to size of one granule, and a little shorter than the medium scale of SBs. On the other hand, if we use the large scale of SBs, 1.1°$\leq D\leq$4.4°, we end up with 0.11°$\leq d\leq$0.7° in good agreement with the medium scale of SBs (Table 1).
In this scenario, $d$ is not the intrinsic size of a granule, but an equlibrium scale resulting from the random walk of flux tubes within a magnetic funnel.

\section{Discussion}

We have studied the geometrical parameters of the chromospheric fibrils lying along a coronal hole boundary in search of solar sources for SBs. The test criteria are the spatial dimension of  SBs (Bale et al 2021, Fargette et al. 2021),  highly elongated shape along the radial direction (Hobury et al. 2002, Laker et al. 2021)  and clustering with a long term memory (Dudok de Wit et al. 2020, Chhiber et al. 2020). With the unprecedented high-resolution H$\alpha$ blue-wing images from GST/VIS of BBSO, we were able to determine those properties of small ejective features down to 100 km scale. 

The chromospheric scales (column 2, 3) measured from the spicules located along the chromospheric networks 
%with the medium and large scales of SBs found from the PSP data (Fargette et al. 2021; Bale et al. 2021).
%The large scale $D$ in the PSP SB patch analysis result is close to a typical network diameter directly measured from the H$\alpha$ images are 1.2$\pm 0.6$° but a little larger (column 3). The inter-distance between any pair of two flux tubes lying along a network boundary is in the range of 0.04°--0.12°.
% smaller than $d$. 
%The inter-distance between neighbor flux tubes shows a population between, which agrees to the medium scale range of SBs (column 2), but it extends to the large scale of SBs, 1.2° with no distinct peaks. 
%Thus, spicules located along the chromospheric networks 
have a little overlap with the medium and large scales of SBs found from the PSP data (Fargette et al. 2021; Bale et al. 2021) but with a tendency of being smaller than those derived from PSP data. The distribution of length and that of inter-distance between every pair of flux tubes also overlap with the residence and waiting time distributions of SBs, but do not show the characteristics of the scale-free single power-law distribution. 
These results raised a couple of issues as to whether a similarity of the SB scales to the supergranulation and granule can evidence solar origin and which else chromospheric parameters can be important for connecting the sun to SBs.

Firstly, the argument for SB modulation by SG does not mean that an SG itself expands to an SB patch maintaining its structure.
%The similarity of the SB scales to those of SG and granule can be directly evidence solar origon if SBs/ patch are SG simply grows into an SB patch. However, 
It means that the open fields lying on the SG boundary lane should expand to form a funnel. Therefore the similarity as evidence for solar origin is justified only when edge of one funnel push agains its neighbor funnel to reproduce their footpoint scale into space. What really maters is then not the original size but the size after the expansion can recover the SG scale. This need is in line with our ad hoc assumption that flux tubes undergo random walk to fill an SG. (Figure 2); otherwise no numerous SBs inside a patch can be detected, and only one or two SBs in the SG funnel will be detected.  
%Likewise, open field located in the granule boundary should also expand to form the medium scale, $d$ of SBs. The medium scale is somewhat larger than actual granule. Expansion (making it thicker) and random walk (repositioning) should have occurred. Otherwise, the flux tubes lined up along the SG boundary with size as thin as an intergranular lane cannot produce quantities comparable to the PSP observation. Our thought experiment suggest that the medium scale should be a result of random walk and the size of its generating driver should be lost. The required rapid expansion of SG lane into SG size should occur near solar corona where magnetic field dominates. To our knowledge, there is no simulation to compare with filamentary structure maintained and random walk grows to make diffusion specifically refering to these two scales.
This hypothesis lead to the relation of the medium scale to the large scale of SBs as $D/d \sim N^{1/2}$.
%, which can be explained with $40\leq N\leq 100$ determined from solar observation. 
Spatial frequency, $N$, is therefore important not only that it connects spicules to SBs, but explains how to link the two scales, $D$ and $d$.

%(2) Occurrence rate (temporal frequency) of solar ejection has been a major issue in search for solar origin. We however suggest that the spatial frequency, i.e., how many flux tubes are packed within a network is also important. If SBs are indeed the spatial distribution (Bale+ 2021), not only that solar sources for SBs should be almost always occuring but the spatial frequency should be high enough to closely match the number of SBs within a patch. For chromospheric SGs we found this number of flux tubes in a so-called magnetic funnels is  40--100 (column 5), which is as high as the large number of SBs in a patch. Neither all of the spicules reach the height of PSP perihelion nor all SBs connected to an SG can be detected by PSP. This quantity also relates the medium scale to large scale of SB as $D/d \sim N^{1/2}$, if the random walk was so rapid in order to make field lines fill up the SG funnel eventually more or less uniformly. Without it, we will be unable to explain the numerous SBs inside a patch. Spatial frequency is therefore an important parameter not only that it directly connects the chromopshere to  with SBs in solar wind, but explains how to link the two scales, $D$ and $d$.

Secondly, we draw attention to the length-to-diameter ratio of the dense section of flux tubes. because aspect ratio of a flux tube can hardly increase during its transit to space in our simple calculation. 
Its value of $6\leq A\leq 40$ (column 4) is already as high as the aspect ratio of SBs ($>$10) measured in space.
%, a property which is otherwise hardly attained during the transport into solar wind. 
Such a high aspect ratio may also allow a sufficiently high probability of individual reversals to be crossed by PSP. An hypothesis that solar flux tubes should develop S-shaped kinks during their transit into the heliosphere is though needed, and our argument is simply that no such structure is found in the sun. Otherwise we refer to the recent finding of no full SBs inside 0.2 au (Pecora et al. 2022) and theoretical models proposing in-situ generation of SBs (e.g., Schwadron \& McComas 2021; Shoda et al. 2021). In this scenario, whether or not twisted structure should form in the sun is unimportant, but whether the aspect ratio and spatial frequency are sufficiently high is important.

Finally we note that we were unable to find a scale-free single power-law distribution from either fluxtube lengths or their inter-distances (column 6). 
This result suggests that the residence and waiting time distributions of SBs are not directly originating from the chromosphere. 
% or alternatively by the self-organized criticality (Aschwanden \& Dudok de Wit 2021). 
We conclude that solar chromosphere may produce seeds for SBs, but solar flux tubes should undergo not only positional rearrangement but also aggregation and cascade to other scales during transit from the sun to solar wind.
%Somefor the number distributions of the length and inter-distance of flux tubes presented in this paper to be related to the waiting-time and residence-time distributions of SBs. 

\acknowledgements
 This work was supported by NASA grants, 80NSSC19K0257 and 80NSSC20K1282, and NSF grant, AGS 2114201. JCMO was supported by NASA grant, 80NSSC19K0257.
 BBSO/GST operation is supported by New Jersey Iinstitute of Technology and US NSF AGS-1821294 grant. GST operation is partly supported by the Korea Astronomy and Space Science Institute and the Seoul National University.

\end{document}